# Ballistocardiogram Signal Processing: A Literature Review

Ibrahim Sadek, *Member, IEEE*

*Abstract*—There are several algorithms for analyzing and interpreting cardiorespiratory signals obtained from in-bed based sensors. In sum, these algorithms can be broadly grouped into three categories: time-domain algorithms, frequency-domain algorithms, and wavelet-domain algorithms. A summary of these algorithms is given below to highlight which category of algorithms will be used in our analysis. First, time-domain algorithms are mainly focused on detecting local maxima or local minima using a moving window, and therefore finding the interval between the dominant J-peaks of ballistocardiogram signal. However, this approach has many limitations because of the nonlinear and nonstationary behavior of the ballistocardiogram signal. The implication is that the ballistocardiogram signal does not display consistent J-peaks, which can usually be the case for overnight, in-home monitoring, particularly with frail elderly. Additionally, its accuracy will be undoubtedly affected by motion artifacts. Second, frequency-domain algorithms do not provide information about interbeat intervals. Nevertheless, they can provide information about heart rate variability. This is usually done by taking the fast Fourier transform or the inverse Fourier transform of the logarithm of the estimated spectrum, i.e., cepstrum of the signal using a sliding window. Thereafter, the dominant frequency is obtained in a particular frequency range. The limit of these algorithms is that the peak in the spectrum may get wider and multiple peaks may appear, which might cause a problem in measuring the vital signs. At last, the objective of wavelet-domain algorithms is to decompose the signal into different components, hence the component which shows an agreement with the vital signs can be selected. In other words, the selected component contains only information about the heart cycles or respiratory cycles, respectively. Interbeat intervals can be found easily by applying a simple peak detector. An empirical mode decomposition is an alternative approach to wavelet decomposition, and it is also a very suitable approach to cope with nonlinear and nonstationary signals such as cardiorespiratory signals. Apart from the above-mentioned algorithms, machine learning approaches have been implemented for measuring heartbeats. However, manual labeling of training data is a restricting property. Furthermore, the training step should be repeated whenever the data collection protocol has been changed.

## I. INTRODUCTION

**B**ALLISTOCARDIOGRAPHY (BCG) is a noninvasive technique for creating a graphical representation of the heartbeat-induced repeated motions of the human body. These repeated motions happen due to the rapid acceleration of blood when it is ejected and moved in the great vessels of the body during periods of relaxation and contraction, known as diastole and systole, respectively. In other words, BCG can provide information about the overall performance of the circulatory system; this is because BCG measures the mass movements, i.e., the mass of the circulating blood and the heart during the cardiac cycle [1]. During atrial systole, when the blood is ejected into the large vessels, the center of mass of the body moves towards the head of the body. In other ways, when the blood moves towards the peripheral vessels and concentrates further away from the heart in the peripheral vessels, the center-of-mass moves towards the feet (Fig. 1(b)). This shift comprises several components as a result of cardiac activity, respiration, and body movements. This shifting of the center of mass of the body generates the BCG waveform since the blood distribution changes during the cardiac cycle [5]. More than 100 years ago, BCG failed to prove its functionality, and it did not start to be used in routine tasks for a few general reasons as follows. First, there had been insufficient standard measurement methods, i.e., different methods had resulted in slightly different signals. Second, the exact physiologic origin of the BCG waveform had not been well-understood. Furthermore, there had been insufficient clear guidelines for interpretation of the results, and therefore the medical community was unwilling to take risks. Third, there had been a dominant focus on some clinical diagnostic, for example, *myocardial infarction*, *angina pectoris*, *coronary heart disease*; these applications need a high level of specificity and reliability that the BCG had not reached. Fourth, the emergence of ultrasound and echocardiography methods that swiftly overhauled BCG and related methods for noninvasive cardiac and hemodynamic diagnostic [6].

At the present time, BCG has been given a lot of interest thanks to the information technology revolution, including hardware technology as well as software and services. BCG sensors can be embedded in ambient environments without the need for medical staff presence. Consequently, it has an outstanding impact in current e-health systems. Ultimately, BCG helps reduce checkups' stress and the patient emotion and attention responses. Fig. 1(a) shows an example of a typical BCG signal, while Fig. 2(b) shows an example of a typical electrocardiogram signal. The BCG waveforms may be grouped into three main groups, i.e., the pre-systolic (frequently disregarded), the systolic and the diastolic as given in Table I. The I and J waves are also quoted as ejection waves [1]. To this extent, the definition, formal limitations, and nomenclature of ballistocardiography were discussed. The formal limitations were mainly due to the complexity of the used system and misinterpretation of the obtained signals and its deformations. The field of ballistocardiography has been revived as a result of the numerous technological advancements, as, for example, the advent of microprocessors and laptop computers. All in all, ballistocardiography can be very useful

I. Sadek is with Image and Pervasive Access Laboratory (IPAL), CNRS UMI 2955, Singapore e-mail: (see ibrahim.sadek@ipal.cnrs.fr).

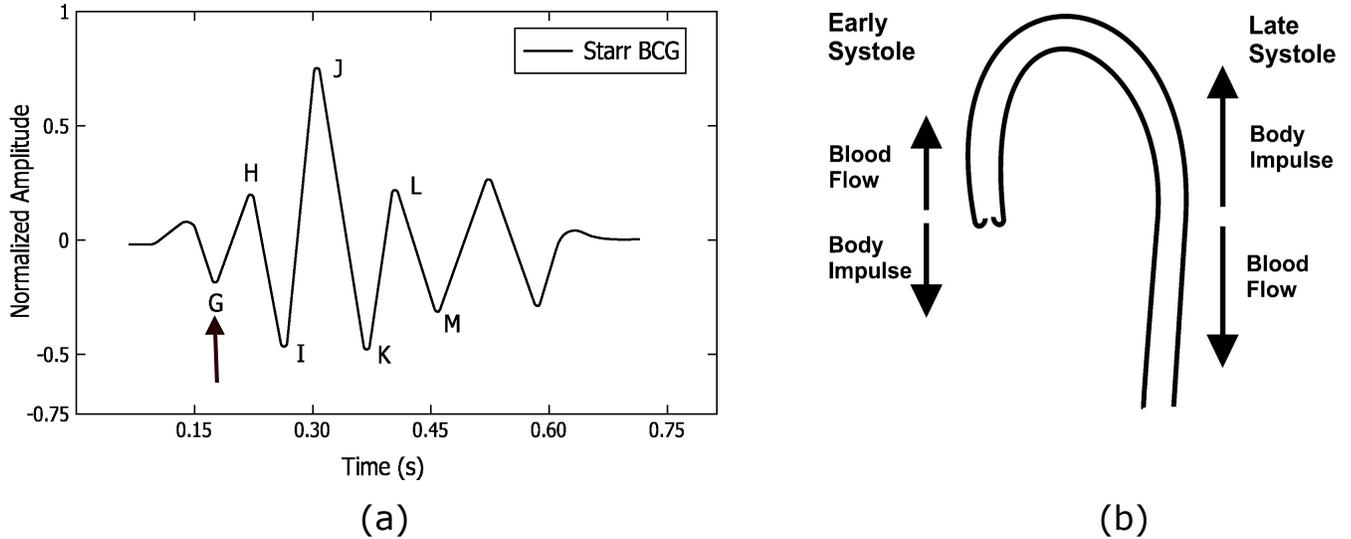

Fig. 1. (a) Example of a typical BCG signal with letters used to designate the waves. The arrow indicates the position of the beginning of the electrical ventricular systole (QRS. complex of the electrocardiogram). Image adapted from [2], [3], [1], (b) Aortic arch and force vectors coming from blood ejection by the left ventricle. Image adapted from [4].

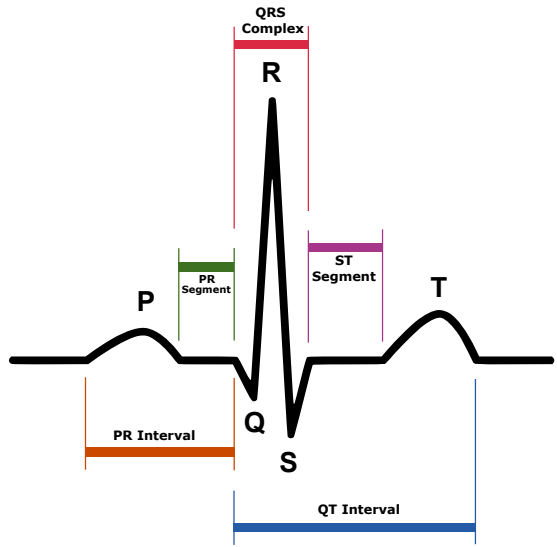

Fig. 2. Example of a typical electrocardiogram signal.

TABLE I
NOMENCLATURE OF BALLISTOCARDIOGRAM (NORMAL DISPLACEMENT) SIGNAL [7], [1].

| *Pre-Systolic Group (See Fig. 1(a))* |
| --- |
| • F wave: (rarely seen) headward wave preceding G, related to pre-systolic events, not an after-vibration. <br> • G wave: small footward wave which at times precedes the H wave. |
| *Systolic Waves (See Fig. 1(a))* |
| • H wave: headward deflection that begins close to the peak of the R wave, maximum peak synchronously or near the start of ejection. <br> • I wave: footward deflection that follows the H wave, occurs early in systole. <br> • J wave: largest headward wave that immediately follows the I wave, occurs late in systole. <br> • K wave: footward wave following J, occurs before the end of systole. |
| *Diastolic Waves (See Fig. 1(a))* |
| • L and N waves: two smaller headward deflections which usually follow K. <br> • M wave: footward deflection between L and N. <br> • Smaller subsequent waves may be visible and are named in sequence. |

in several applications such as monitoring of cardiac function and performance in addition to monitoring of sleep and sleep-disordered breathing [8], [9]. One of the most prominent features of ballistocardiography is the accessibility and ready-availability, which allows the system to be deployed in users' homes without affecting the users' privacy and daily activities. In what follows, we explain more in detail the various tools and algorithms exist in the literature to analyze and interpret ballistocardiography, wherein we look at what types of sensors that can be used for signal acquisition and what types of software algorithms that can be used to extract vital information such as heartbeat, respiration, and body movements.

## II. PIEZOELECTRIC POLYVINYLIDENE FLUORIDE-BASED SENSORS

The piezoelectric effect is the ability of some materials to produce an electric charge in response to applied mechanical stress. The polyvinylidene fluoride (PVDF) is an exciting piezoelectric material and is usually developed as a very thin and easily bent film. If a pressure force is applied to the film, it creates a mechanical bending and a shifting of positive and negative charge centers in the film, which then results in an external electrical field. The charge generated from PVDF is equivalent to the applied pressure. Therefore, PVDF is one of the suitable candidates for detecting the small fluctuations generated by different body parts [10].

Wang *et al.* [11] proposed to use a PVDF piezopolymer film sensor for unconstrained detection of respiration rhythm and pulse rate. The film sensor was placed under the bedsheet at the location of the thorax to obtain the variations of the pressure on the bed attributable to respiratory movement and heartbeats. The authors used the wavelet multiresolution decomposition to compute the respiration and heartbeat. The output of the respiratory inductance plethysmography (RIP) and electrocardiography (ECG) were used as a reference for respiration and heartbeat, respectively. The objective of the wavelet analysis was to decompose the raw signal into low-frequency components and high-frequency components. Next, the component presenting a good agreement with either the respiratory movement or the heartbeat was selected. Afterward, the respiratory rate was computed directly based on a time-varying adaptive threshold. On the other hand, the heartbeat component was first squared to rectify it into unipolar, and then the envelope of the rectified signal was calculated using a moving average smoothing algorithm. At last, a time-varying adaptive threshold was also applied to the smoothed envelope to compute the heart rate. It should be noted that heart rate detection was very challenging because the pressure variations attributable to heartbeat on the bed was very weak, and the shape of the signal was not always uniform. Another study was proposed by Wang et al. [12] to detect respiration rhythm and pulse rate of premature infants using PVDF sensor array. The system was tested in clinical environments on five premature infants (1 male and 4 females). The main challenge of the proposed system was frequent body movement of the infants and the weakness of the heartbeat vibration.

Niizeki et al. [13] suggested using a PVDF sensor array for unconstrained monitoring of respiration and heart rate. The sensor array consisted of eight PVDF cable sensors and they were horizontally integrated with a textile sheet on a bed surface covering the upper half of the body. The cardiorespiratory signals, i.e., BCG and respiration were obtained using infinite impulse response digital filters. After extracting the cardiorespiratory signals, an optimal sensor selection search routine was applied to select the most appropriate sensor. The selection criterion was based on the magnitude of the power spectrum density (PSD). The autocorrelation functions of the cardiac and respiratory signals were computed using a 5-second and 15-second time segments for heartbeat and respiration, respectively. The outputs of the autocorrelation functions were smoothed and differentiated using a Savitzky-Golay (5 adjacent points) algorithm and finally, the heart rate and respiration were computed by measuring the intervals between the peaks for the respective autocorrelation functions. A fixed threshold was used to determine if the subject changes posture during the measurement, in which the output from the PVDF cables was disturbed to a large extent. A charge-coupled device (CCD) camera was used to record the image of the body position during posture change as a time stamp. The proposed system was tested against thirteen healthy male subjects whose ages ranged from 21 to 49 years. ECG and pneumotachometer for measuring respiratory flow were used as a reference during the study. The study consisted of two phases, i.e., short-term recording for 10 minutes and an overnight study for 2 hours. For the overnight recording, only 7 subjects were involved. The proposed system had some limitations in particular susceptibility to motion artifacts caused by subject movements that might have led to the misidentification of the peak for autocorrelation functions.

Paalasmaa and Ranta [14] applied an unsupervised learning approach on ballistocardiogram signals to compute heartbeat. The ballistocardiogram signals were collected from three subjects using a piezoelectric pressure sensor over 5 hours recording. To start with, feature vectors were extracted from the signal at possible heartbeat positions, i.e., the local maxima of the signal. Then, a complete-link clustering was applied to the feature vectors to look for a cluster with the highest density. The positions of the feature vectors of the densest cluster were found to match real heartbeat positions in the signal. An angular dissimilarity measure was adopted since it omits the differences in feature vector amplitudes. The sensor was located close to the patient's upper body so that it can register cardiac activity in a proper way.

Paalasmaa et al. [15] introduced a sleep tracking web application, which was based on measurements from a piezoelectric film sensor placed under the mattress topper. The raw data coming from the sensor was sent to a web server for analysis and extracting information. This information includes heart rate, respiration, sleep staging, and stress reactions. The heart rate was computed by creating a heartbeat template using complete-link clustering [14], then the heart rate intervals were detected by selecting those intervals that minimize a predetermined residual error. The sleep staging was carried out by utilizing heart rate variation, respiration variation, and activity information. The proposed approach was validated against a 40-patient group at a sleep clinic. The added value of this work is the suitability of the system for long-term monitoring of sleep and the web application for sleep analysis at home. A more comprehensive study was introduced by Paalasmaa et al. [16] to compute heart rate from ballistocardiogram signals acquired with piezoelectric film sensor. At first, a model for the heartbeat shape was adaptively deduced from the signal using a hierarchical clustering approach. Afterward, interbeat intervals were identified by detecting positions where the heartbeat shape best matches the signal. The proposed method was verified with overnight recordings from 46 subjects in different settings, i.e., sleep clinic, home, single bed, and double bed.

Chen et al. [17] advised to use four piezoelectric sensors to detect heart rate and respiration. One sensor was placed under the pillow, whereas the other three were placed under the mattress close to the back, hip, and calf level positions. The data was collected from five healthy subjects at age of twenties during a 2-hour's nap in a sleep lab. ECG and nasal thermistor signal were employed as heart rate and respiration references. Heart rate and respiration were computed based on the multiresolution analysis of the wavelet decomposition in which the Cohen–Daubechies–Feauveau biorthogonal wavelet was selected as the basis function to design the decomposition and reconstruction filters. The 6[th] level approximation waveform was similar to the respiratory rhythm, while a combination of the 4[th] and 5[th] scale coefficients were found to be suitable for heart rate detection. The authors were able to measure both vital signs from the four positions. However, the overall optimal position was found in the back. That makes sense because the more the sensor is closer to the thorax, the more accurate the recovered signals are.

A wheelchair-based system for monitoring the cardiac activity of its user was proposed by Pinheiro *et al*. [18]. The signals were collected from piezoelectric film sensors and micro-electromechanical systems accelerometers installed in the seat and backrest of the chair. The system also included photoplethysmography (PPG) sensors in the armrests. The data from the sensors were sent via Wi-Fi to a laptop with a data acquisition board for deeper analysis. ECG recordings were used to validate the proposed system. The system was tested in different situations, namely unmoving wheelchair, tiled floor motion, and treadmill tests. In the last two situations, the ballistocardiogram signals collected from the piezoelectric sensors were completely corrupted by motion artifacts. On the other hand, the accelerometer was much more insensitive to wheelchair motion. The analysis was done on seven subjects using the fast Fourier transform. Subsequently, the prominent peak was selected within a specific frequency range for heart rate estimation. In a summary, getting informative ballistocardiogram signals from the piezoelectric sensors in a motion situation was almost impossible. However, it was more convenient to get informative signals from the accelerometers and the PPG sensors.

A multichannel approach was proposed by Kortelainen *et al*. [19] to extract heart rate and respiration information using eight PVDF sensor channels located in the upper position of the bed. The heart rate was estimated by averaging the signal channels in the frequency domain, in which a sliding time window was utilized to compute the cepstrum of each signal channel. However, the respiratory rate was computed from the first principal component of a principal component analysis (PCA) model applied to the low-pass filtered bed sensor signal. The assumption was that the first principal component will give the signal with the maximum variance, and as a result shall improve the sensitivity for the extraction of the respiration. Twenty-eight patients were recruited for the study and they were suspected to have diverse kinds of sleep problems. Frequency domain averaging was better than simple averaging over all the sensor channels. The extracted information, i.e., heart rate, respiration, and movement might have been used for further sleep analysis.

The same pressure bed sensor assembly with eight PVDF sensors was applied for sleep apnea detection in [20]. The respiratory signal was computed by two methods. The first method was to apply a Hilbert transform to the bed sensor signal and then smooth the signal with a low pass filter. The second method was similar to Kortelainen *et al*. [19] by adopting the PCA approach. At last, the amplitude baseline of the respiratory signal was estimated as the mean value of the preceding 100 seconds. An apnea event was detected if the ratio with the baseline was less than a selected percentage threshold value for a period of at least 10 seconds. The authors applied their methodology to twenty-five patients out of twenty-eight patients recruited in [19]. The system showed a good agreement with the reference polysomnography. However, the authors used the simplified reduced respiratory amplitude index (RRAI) instead of the standard apnea-hypopnea index (AHI). In another study, Brüser *et al*. [21] have implemented three different methods using the same sensor set to measure the heart rate in a nonintrusive way. Initially, the heart rate was computed using a sliding window cepstrum analysis [19]. Secondly, the heart rate was computed using a Bayesian fusion approach, in which three estimators were calculated from each sensor channel such as adaptive-window autocorrelation, adaptive-window average magnitude difference function, and maximum amplitude pairs. For each channel, these three estimator outputs were then combined using a Bayesian fusion method to obtain an overall estimate. In other words, Bayesian fusion approach was applied to 24 estimates. At last, the heart was estimated based on the aforementioned approach. However, for each channel separately. In general, the multichannel based approaches improved the robustness of heartbeat interval estimation over a single sensor. More specifically, Bayesian-based method slightly outperformed the cepstrum-based method.

Martin-Yebra *et al*. [22] extracted heart rate variability indices from ballistocardiogram signals and then evaluated their correlation with electrocardiogram-derived ones. The ballistocardiogram signals were acquired by a piezoelectric 3D-force plate in supine and standing positions, in a group of 18 healthy subjects (11 females). For each position, the data collection was performed during 5 minutes. Furthermore, subjects were asked to stay quiet to avoid any motion artifacts. The ballistocardiogram waves, i.e., (H, I, J, K) were detected by synchronizing ballistocardiogram signals with ECG signals. Although the proposed approach provided a good match with the reference ECG, it is very difficult to generalize this approach for real-life deployment as the data collection was conducted for a very short time and the detection part was achieved by adapting information from the ECG signals.

Katz *et al*. [23] measured cardiac interbeat intervals using a contact-free piezoelectric sensor placed beneath the mattress under the tested subjects. The data was collected from 25 home sleep recordings of 14 healthy subjects in a two-in-bed setting. The authors applied three algorithms to the collected ballistocardiogram signals as follows. First, interbeat intervals were found by decomposing the signal into multiple components using an empirical mode decomposition filter and then locating the candidate peaks within a localized search area. Second, after locating potential interbeat intervals, a binomial logistic regression model was applied to classify each interbeat interval into one out of three groups based on morphological properties of the ballistocardiogram signal. Finally, an additional algorithm was implemented to get discrete interbeat interval distribution maps during the night recording, considering interbeat interval data from overlapping 15 minutes windows. The preceding three algorithms demonstrated the effectiveness of the proposed system for heart rate variability analysis. Sela *et al*. [24] used the same piezoelectric sensor to detect left ventricular ejection for 10 subjects (6 males and 4 females), where the lower body of each subject was enclosed in a negative pressure chamber. The negative pressure chamber regulates and controls the blood pressure of the participants. This study demonstrated the ability of the system to identify internal bleeding condition among patients at risk, namely individuals after an accident or surgical operation.

Alvarado-Serrano et al. [25] measured beat-to-beat heart rate from subjects sitting in a common office chair. The authors used a piezoelectric sensor fixed to the bottom side of the seat to collect ballistocardiogram signals from seven subjects (5 males and 2 females). Continuous wavelet transform with splines was implemented to detect beat-to-beat intervals in which an optimal scale was selected to reduce noise and mechanical interferences. Thenceforth, learning and decision phases where applied to the selected scale to detect potential J-peaks. In the learning phase, the first four heartbeats in the ballistocardiogram signal were found to define initial thresholds, search windows, and interval limits. The learned parameters were then utilized to determine the next heartbeat and were readopted after each heartbeat detected to adhere to the heart rate and signal-amplitude changes. A similar study was proposed by Liu et al. [26]. However, two PVDF film sensors were installed in the seat cushion and foot insole.

Choe and Cho [27] used a piezoelectric sensor installed between a bed-frame and a mattress for unconstrained monitoring of heart rate. The data was collected from 7 male subjects sleeping in a supine sleeping position where the sensor was placed under the subject's back. In total, they collected ballistocardiogram signals for about 5 hours from all subjects, in which subjects were not moving during data acquisition. The data was first smoothed using a moving mean absolute deviation, then the J-peaks were detected within a specific search region using an adaptive thresholding technique. The authors achieved satisfactory results with the reference ECG. However, this method may not be applicable in real-life applications because the data was not collected in a typical sleep sitting and the motion artifacts were not considered as well. Table II summarizes the unconstrained monitoring of vital signs using the PVDF-based sensors.

### III. Electromechanical Film-Based Sensors

The electromechanical film (EMFi) material is a plastic film that can transform mechanical energy into an electrical signal and the other way around. Basically, it is a flexible and thin bi-axially oriented polypropylene film covered with electrically conductive layers, which are enduringly polarized. EMFi has a static charge reaching hundreds of Volts. When a pressure is applied to the film, a charge is created on its electrically conductive surfaces and this charge can be measured as a current or voltage signal, usually with a charge amplifier. As a result, the EMFi serves as a sensitive motion sensor [28]. Alametsä et al. [28] suggested to use EMFi sensors for obtaining ballistocardiogram signals from certain places of the body. The authors installed EMFi sensors in a chair and in smaller pieces in a few positions on the body (arm, leg, and chest). The ballistocardiogram signals were collected from a few people and the duration of the recordings was relatively short. This study demonstrated the potential of the EMFi material in monitoring the changes in cardiac function. In another study, Koivistoinen et al. [29] evaluated the ability of the EMFi sensors for measuring ballistocardiogram signals. The authors installed two EMFi sensors in the seat and backrest of a normal chair, and the data was collected from two young subjects (1 male and 1 female) for 5 minutes. After visual inspection versus the reference ECG, it was found that the acquired waveforms closely simulate those reported in the literature. Equivalent results were also reported by Junnila et al. [30], [31], which presented the suitability of the EMFi sensors for extracting ballistocardiogram signals.

A smart mattress was developed by Koivistoinen et al [32] to detect interbeat intervals in a nonintrusive way from six male subjects. The mattress consisted of 160 EMFi electrodes distributed throughout the mattress that enabled signal acquisitions from multiple locations. Two methods were implemented to detect interbeat intervals, i.e., a pulse method and an adaptive window cepstrum method. In the former, signals from all channel were high pass filtered and then squared. After that, these squared values were averaged between all channels and low-pass filtered the result. At last, the beginning of each heart rate was tracked in the generate pulse train signal. In the latter, the window length of the cepstrum was selected using the pulse method as the first estimator of the heart beats. Then, signals from all channels were averaged in the frequency domain. An interpolation was used to detect more accurate location for the selected cepstrum maximum value. Moreover, the motion artifacts were eliminated based on the signal variance using a sliding time window. Although the cepstrum-based method provided better results than the pulse method, its computational efficiency was not as good as the adaptive window method.

Aubert et al. [33] adopted a single EMFi sensor to provide heart rate, breathing, and an activity index representing body movements. The recommended system was validated utilizing data collected from 160 subjects (58 males and 102 females) for a total of 740 hours. Part of the data was collected in a sleep laboratory from patients (i.e., sleep apnea, insomnia, and other sleep disorders) who underwent a full polysomnography and the other part was collected at home from healthy subjects. Body movements were first isolated from the sensor data based on the signal amplitudes and energy, and their time derivatives. Thereafter, heart rate was measured using a sliding window autocorrelation method, in which the optimal window length had to span 3 to 5 consecutive beats. The respiratory rate was estimated based on the local peaks, troughs, and zero-crossings, constrained to rules ensuring physiological validity in terms of duration and amplitude. Across the 60 subjects, the vital signs were computed over epochs of 30 seconds and the average values were computed and compared to the reference ECG and thorax belt, respectively. The recommended system achieved satisfying results compared to the reference devices.

Kärki and Lekkala [34] used EMFi and PVDF sensors in the measurements of heart rate and respiration. The objective of the study was to determine if there were differences between the results of both sensors. ECG was used as a reference for heart rate and a thermistor for respiration rate. Heart rate and respiration were measured using power spectral density (PSD). The two sensors were embedded inside a textile pocket and the pocket itself was integrated into clothing. They were positioned underneath a commercial heart rate belt on the left side of the sternum. Preliminary results showed that both sensors provided reliable results in the measurements of heart and respiration rates. However, the PSD was not robust enough because the peak in the spectrum might get wider and multiple peaks might have appeared. Another study was proposed by Kärki and Lekkala [35] to determine heart rate with EMFi and PVDF materials. The EMFi and PVDF sensors were grouped together to a form a single structure. The data was collected from 10 subjects (5 males and 5 females) over 60 seconds recording (sitting and supine positions), where the sensor structure was placed under the legs of a chair and bed. These preliminary results demonstrated that the heart rate can be measured at home just by sitting on a chair or lying in a bed.

TABLE II
SUMMARY OF UNCONSTRAINED MONITORING OF VITAL SIGNS USING PVDF-BASED SENSORS. *WT*: WAVELET TRANSFORM, *N/A*: NOT AVAILABLE, *P. Infants*: PREMATURE INFANTS, *M*: MALE, *F*: FEMALE, *HR*: HEART RATE, *HRV*: HEART RATE VARIABILITY, *RR*: RESPIRATORY RATE, *ACF*: AUTOCORRELATION FUNCTION, *Min*: MINUTES, *Hrs*: HOURS, *Sec*: SECONDS, *CLC*: COMPLETE-LINKAGE CLUSTERING, *TM*: TEMPLATE MATCHING, *FREQ*: FREQUENCY, *CEP*: CEPSTRUM, *PCA*: PRINCIPAL COMPONENT ANALYSIS, *MAP*: MAXIMUM AMPLITUDE PAIRS, *AMDF*: ADAPTIVE-WINDOW AVERAGE MAGNITUDE DIFFERENCE FUNCTION, *ECG Sync*: ELECTROCARDIOGRAM SYNCHRONIZATION, *EMD*: EMPIRICAL MODE DECOMPOSITION, *TH*: THRESHOLD, *CWT*: CONTINUOUS WAVELET TRANSFORM, *Lab*: LABORATORY.

| | Method | Subjects (M, F) | Deployment | Duration | Outcome |
|---|---|---|---|---|---|
| [11] | WT | N/A | Lab | N/A | HR, RR |
| [12] | WT | 5 P. Infants (2 M and 3 F) | Hospital | 10 Min | HR, RR |
| [13] | ACF | 13 M | Home | 10 Min, 2 Hrs | HR, RR |
| [14] | CLC | 3 N/A | Lab | 330 Min | HR |
| [15] | CLC, TM | 40 N/A | Sleep clinic | Overnight | HR, RR |
| [16] | CLC, TM | 60 N/A | Sleep clinic, home | Overnight | HR |
| [17] | WT | 5 N/A | Lab | 2 Hrs | HR, RR |
| [18] | FREQ | 21 N/A | Wheelchair | 5 Min | HR |
| [19] | CEP, PCA | 6 N/A, 15 M, 13 F | Hospital | Overnight | HR, RR |
| [20] | PCA | 15 M, 13 F | Hospital | Overnight | Apneas |
| [21] | ACF, MAP, AMDF | 15 M, 13 F | Hospital | Overnight | HR |
| [22] | ECG Sync | 17 M, 11 F | Lab | 5 Min | HRV |
| [23] | EMD | 14 N/A | Home | Overnight | HR |
| [24] | N/A | 6 M, 4 F | Lab | 84 Min | LVET |
| [25] | CWT | 5M, 2 F | Chair | 100 Sec | HR |
| [26] | Adaptive TH | 7 M | Lab | 45 Min | HR |
| [27] | CWT | 6 N/A | Lab | 67 Min | HR |

Pinheiro et al. [36] introduced a low-cost system to measure blood pressure variability and heart rate variability. A single EMFi sensor was installed in the seat of a normal office chair to measure ballistocardiogram signals while a finger PPG was used to estimate arterial oxygen saturation (SpO2). For validation, ECG was acquired using three chest leads. Using LabVIEW, heart rate and heart rate variability were determined by an adaptive peak detection algorithm. The pulse arrival time was estimated as the time difference between ECG and PPG maximum peaks, and when considering BCG-PPG relation, the I-valley (Fig. 1(a)) was the reference. The designed system was appraised using data collected from five healthy volunteers over 10 minutes recording. The preliminary study demonstrated that heart rate variability can be measured using the correlation between BCG and PPG. The PSD was exploited to measure the heart rate. In another study, Pinheiro et al. [37] collected ballistocardiogram signals by placing an EMFi sensor in the backrest of a wheelchair's, beneath the lining. Two modulation-based schemes were carried out for heart estimation, i.e., a sliding power window and an all-peak detector. The objective was to find all local maxima and local minima, then a spline interpolation and a moving power window were employed to compute a modulating signal. At last, a fast Fourier transform was applied to the output of each method in order to measure the average heart rate from the signal's fundamental frequency. This system was evaluated using data gathered from six normal subjects (4 males and 2 females) during 125 seconds.

Brüser et al. [38] proposed an unsupervised approach to determine inter-beat intervals using an EMFi sensor. The sensor was fixed underside of a thin foam overlay which was thus located on top of the mattress of a typical bed. The system was evaluated on over-night recordings from 33 individuals (14 males and 19 females). Three estimators were implemented, namely autocorrelation function, average magnitude difference function, Maximum amplitude pairs in order to compute the local interval length using a sliding time window. Ideally, this window contained two events of interest. The values of the local interval length were constrained by two thresholds, i.e., $T_{min}$ and $T_{max}$. The body movements were detected based on the maximum amplitude range of each time-window. The information from the three estimators was then applied to a probabilistic Bayesian method to estimate the inter-beat intervals in a continuous manner. Although the proposed method achieved very satisfactory results, the main limitation existed in the implicit hypothesis that two successive heart beats in the BCG have an unknown but similar morphology. This assumption may not always hold true.

In the same way, Zink et al. [39] used an EMFi sensor to detect heartbeat cycle length in patients suffered from atrial fibrillation and sinus rhythm. The sensor was placed under the bed-sheet and data was collected from 22 patients (15 M, 7 F) during and after cardioversion. Cardioversion is a medical procedure that returns a normal heart rhythm in people with certain types of abnormal heartbeats, namely arrhythmias. In another study, Zink et al. [40] employed the EMFi sensor to measure heartbeat in patients suffered from sleep-disordered breathing. Twenty-one patients (19 males, 2 females) were recruited for the study and underwent a standard full-night polysomnography. A quality-index was proposed based on the three estimators previously discussed in [38] that allowed to identify segments with artifacts and to automatically exclude them from the analysis. The proposed system provided good correlation of beat-to-beat cycle length detection with simultaneously recorded ECG.

Pino et al. [41] used two EMFi sensors installed in the seat and backrest of a normal chair in order to measure heart rate. Ballistocardiogram data were collected from 54 individuals, whereas 19 of them were measured in a laboratory (1 minute) and the rest in a hospital waiting room (2 minutes). Firstly, empirical mode decomposition and wavelet analysis were (Deabuchie 6) implemented to reconstruct ballistocardiogram signal. Secondly, the J-peaks of the ballistocardiogram signal was detected using a length transform analysis. The body movements were eliminated using a moving time window. Then, for each time-window two thresholds were computed, i.e., $T1 = (max + min)/2$ and $T2 = mean + 1.1 * std$, if T1 was greater than T2, the current window was marked as a body movement. The wavelet analysis was preferred to reconstruct the signal as it produced a higher effective measurement time. A similar approach was also proposed by Pino et al. [42]. However, they increased the size of the dataset to 114 people. Of those, 21 were gathered in a school (2 minutes), 42 in homes (2 minutes), and 51 in a hospital waiting area. It is difficult to assess the robustness of this system because the data was collected in a very short time and in a controlled environment as well.

In a recent study, Alametsä and Viik [43] presented the stability of ballistocardiogram signal during 12 years' time, on which the data was gathered from a single person in a sitting position using EMFi sensors. Several other signals were recorded as well such as ECG, ankle pulse signal, and the carotid pulse signal from the neck near the carotid artery. All measurements lasted about 2 to 3 minutes with a sampling frequency of 500 Hz. In a conclusion, ballistocardiogram research may be recommended for examining long-term changes in heart operation and to reveal variations in it. Table III summarizes the unconstrained monitoring of vital signs using the EMFi-based sensors.

## IV. PNEUMATIC-BASED SENSORS

The idea of the pneumatic system is to deploy a thin air-sealed cushion between the bed and mattress. Thereafter, when a person rests in the bed, the forces originated because of the heartbeat, respiration, snoring and body movements affects the air in the cushion through the mattress. This slight human movement causes a pressure and therefore variations in pressure are measured by a supersensitive pressure sensor [44], [45].

TABLE III
SUMMARY OF UNCONSTRAINED MONITORING OF VITAL SIGNS USING EMFI-BASED SENSORS. *WT*: WAVELET TRANSFORM, *N/A*: NOT AVAILABLE, *M*: MALE, *F*: FEMALE, *HR*: HEART RATE, *RR*: RESPIRATORY RATE, *ACF*: AUTOCORRELATION FUNCTION, *Min*: MINUTES, *Hrs*: HOURS, *Sec*: SECONDS, *CEP*: CEPSTRUM, *MAP*: MAXIMUM AMPLITUDE PAIRS, *AMDF*: ADAPTIVE-WINDOW AVERAGE MAGNITUDE DIFFERENCE FUNCTION, *EMD*: EMPIRICAL MODE DECOMPOSITION, *TH*: THRESHOLD, *LT*: LINEAR TRANSFORM, *Lab*: LABORATORY.

|      | Method            | Subjects (M, F) | Deployment      | Duration      | Outcome |
|------|-------------------|-----------------|-----------------|---------------|---------|
| [32] | Visually          | 1 M, 1 F        | Lab             | 5 Min         | BCG     |
| [32] | CEP               | 6 M             | Lab             | Overnight     | HR      |
| [33] | Adaptive TH, ACF  | 58 M, 102 F     | sleep Lab, Home | Overnight     | HR, RR  |
| [34] | PSD               | N/A             | Lab             | 60 Sec        | HR, RR  |
| [35] | PSD               | 5 M, 5 F        | Lab             | 30 Sec        | HR, RR  |
| [36] | PSD               | 5 N/A           | Lab             | 10 Min        | HR, BP  |
| [37] | PSD               | 4 M, 2 F        | Lab             | 125 Sec       | HR      |
| [38] | ACF, MAP, AMDF    | 14 M, 19 F      | Clinic          | Overnight     | HR      |
| [39] | ACF, MAP, AMDF    | 15 M, 7 F       | Hospital        | N/A           | HRV     |
| [40] | ACF, MAP, AMDF    | 19 M, 2 F       | Hospital        | Overnight     | HR      |
| [41] | EMD, WA, LT       | 54 N/A          | Lab, Hospital   | 1 Min, 2 Min  | HR      |
| [42] | EMD, WA, LT       | 114 N/A         | Home, Hospital  | 2 Min, 2 Min  | HR      |

Watanabe *et al.* [46] used the aforementioned pneumatic system to measure heartbeat, respiration, snoring, and body movements in a noninvasive manner. The three bio-signals, namely heartbeat, respiration, and snoring were detected using a band-pass filter with different cutoff frequencies. Following, windowed Fast Fourier transform algorithm was applied to measure heart rate and respiration. However, the relative magnitude of snoring was calculated by the standard deviation of the filtered snoring signal and the relative magnitude of body movements was calculated as the standard deviation of the envelope of the sensor output signal. The authors validated the proposed system using data collected from 15 subjects (12 males and 3 females) over 15 nights. Preliminary results showed good agreement against reference devices, namely ECG, belt-type respirometer, and a snoring detection microphone. The body movements were identified and recorded by a CCD camera. In another study, Kurihara and Watanabe [47] acquired data from 10 subjects (20 seconds each) to measure heart rate and respiration. In this study, a condenser microphone was used as a reference for heart rate, respiration and signal-to-noise ratio. Validation results demonstrated that the pneumatic system was more susceptible to environmental noise, for example, opening and closing the door than the reference condenser microphone.

Chee *et al.* [48], [49] recommended to use a balancing tube between two air cells to improve the effectiveness of posture changes during data collection. Balancing tube with a high air resistance aimed at equalizing the pressure of each air cell within a certain time constant. More precisely, it performed the role of a high-pass filter to eliminate body motion. The air-mattress system consisted of 19 air cells, in which measurements can be performed between any pair of cells. However, the authors collected data from the two cells situated on the backside of the chest and abdominal region. Signal was collected from a single subject laying on the air mattress where ECG and nasal airflow signal were collected simultaneously. Although the balancing tube helped eliminate body motion, it affected the sensitivity of the measurement. Heart rate was measured by finding the maximum peak of the BCG signal between the two R-R peaks of the ECG signal. On the other hand, the respiratory rate was measured by windowed fast Fourier transform, i.e., short-time Fourier transform (STFT). Preliminary results showed good match against reference devices. Nevertheless, the proposed system might not be a preferred choice for large-scale deployment due to its complexity. In another study, Shin *et al.* [50] applied the same air mattress for uncontaminated measurement of heart rate and respiration. In which, a total of 13 healthy male subjects were involved in the validation study, i.e., four hours study. The authors measured the heart rate from the R-peaks of the ECG, while the respiratory rate was measured manually. In addition, the authors asked three subjects to simulate sleep apnea (breath-holding) five times each for 10 to 15 seconds. Thereafter, the apneas were detected based on the variance of the respiratory signal with a moving window technique. Table IV summarizes the unconstrained monitoring of vital signs using the pneumatic-based sensors.

## V. STRAIN GAUGES-BASED SENSORS

Brink *et al.* [51] implemented four force sensors under bed-frames to unobtrusively record heartbeat, respiration activity, and body movements. Each force sensor consisted of a reflex light barrier sandwiched between two aluminum plates. When a force is applied to the sensor, the two aluminum plates are squeezed together slightly and the distance between them decreases. The reflex light barrier senses the distance between the two plates and converts it into a voltage signal, which is analogous to the ballistic forces of the heart. This voltage signal is then pre-amplified and passed through a low-pass filter to eliminate ripple and noise. In this preliminary study, heartbeat and respiration were detected by finding local minima or maxima in the signal within a sliding window. To evaluate the robustness of the force sensors, the signals were acquired from four subjects (2 males and 2 females) and in different conditions, i.e., three types of single beds, three types of frames, two types of mattresses. In total, seventy-two conditions were evaluated. In each condition, subjects were asked to sleep in a relaxed supine position on the bed. The signals were collected during 5-minute recording from the four force sensors. Additionally, ECG signals were also collected as a reference. Preliminary results showed that the proposed system can be an acceptable tool for computerized and unattended sleep-data collection over a lengthy period.

Inan *et al.* [52] collected ballistocardiogram signals using strain gauges within a modified commercial scale. The signals were collected from twenty-one subjects (11 males and 10 females), on which participants were asked to stand as quiet as possible on the scale for 45 seconds while BCG and ECG were concurrently recorded. In this study, the measured ballistocardiogram signals from all subjects closely resemble those reported in the literature. Besides, the system was able to provide beat-to-beat cardiac output monitoring. Additionally, ballistocardiogram measurements were found to be repeatable over 50 recordings collected from the same subject over a three-week period. The proposed solution was more susceptible to motion artifacts because the signals were acquired in a standing position. Hence, it might not be suitable for older adults who cannot stand as tranquil on the scale. In order to eliminate floor vibrations, Inan *et al.* [53] proposed a seismic sensor, i.e., geophone, located in proximity to the modified scale that served as the noise reference. An adaptive algorithm was then implemented to filter the output of this sensor and cancel the vibrations from the measured ballistocardiogram signal. Signals were collected from a healthy volunteer while another person stomped around the scale, hence producing increased floor vibrations. Furthermore, signals were also collected from another volunteer standing inside a parked bus while the engine was functioning. This research established that ballistocardiogram recording is feasible in almost all environments, including ambulances and other transport vehicles, as long as the vibrations are not so significant to rail the electronics or lead to a distorted version of the ballistocardiogram force to be coupled to the scale.

TABLE IV
SUMMARY OF UNCONSTRAINED MONITORING OF VITAL SIGNS USING PNEUMATIC-BASED SENSORS. *N/A*: NOT AVAILABLE, *M*: MALE, *F*: FEMALE, *HR*: HEART RATE, *RR*: RESPIRATORY RATE, *Min*: MINUTES, *Hrs*: HOURS, *Sec*: SECONDS, *STFT*: SHORT-TIME FOURIER TRANSFORM, *Lab*: LABORATORY.

|  | Method | Subjects (M, F) | Deployment | Duration | Outcome |
|---|---|---|---|---|---|
| [46] | STFT | 12 M, 3 F | Lab | Overnight | HR, RR, SI |
| [47] | STFT | 10 N/A | Lab | 20 Sec | HR, RR |
| [48], [49] | ECG Sync, STFT | 1 N/A | Lab | N/A | HR, RR |
| [50] | ECG Sync, STFT | 13 M | Lab | 4 Hrs | HR, RR |

In the same way, Inan *et al.* [54] evaluated the electromyogram signal collected from the feet of the subject during ballistocardiogram recording as a noise reference for standing ballistocardiogram measurements. As the lower-body electromyogram signal can be collected directly from the footpad of the modified scale, the proposed system is self-contained and can automatically eliminate motion artifacts. In another study, Wiard et al. [55] used a motion sensor instead of electromyogram sensors to record body motions and to serve as a noise reference. The added value of the motion sensor was to provide a minimum delay between the motion-related noise in the measured signal and the noise detected by the motion sensor. This minimum delay provided the time resolution needed to flag single heartbeat events, hence maximizing the refinement of the approach.

Brüser *et al.* [56] introduced an unsupervised learning approach to measuring heartbeat in a noninvasive manner. Ballistocardiogram signals were recorded by strain gauges in a Wheatstone bridge configuration attached to the slat under the mattress of a hospital bed. A high-pass filter was applied to the raw data in order to remove low-frequency respiratory components. Next, a set of features, representing the fundamental morphology of the heartbeat, were extracted from a 30-second time segment. Afterward, the principal component analysis was applied in order to reduce the dimensionality of the feature vectors. Additionally, a k-means clustering algorithm was adopted to identify clusters of feature vectors. This training step resulted in a list of estimated heartbeat locations. The parameters obtained during the training step were thus manipulated to locate heartbeats in the remaining ballistocardiogram signal by merging the results of three independent indicator functions, i.e., cross-correlation, Euclidean distance, and heart valve signal. Finally, the estimated heartbeat locations were exploited to provide an improved list of beat-to-beat periods. Signals were captured from sixteen healthy subjects (9 males and 7 females) during thirty minutes switching their positions every 7.5 minutes (left lateral, supine, right lateral, prone). This method produced good agreement with the reference ECG. However, the primary limitation was the training step as it had to be repeated whenever subjects enter the bed or adjust their posture with regard to the ballistocardiogram sensor.

Nukaya *et al.* [57] provided a contact-free method for unobtrusive measuring of heartbeat, respiration, body movement, and position change. The authors collected the pressure data using four piezoceramics transducers set beneath bed supports. The proposed system was able to detect previous bio-signals without the need for a preamplifier, accordingly without any voltage source. This is because the sensing devices were distortion sensors that operate without an electrical power supply, i.e., they produce voltage according to the time derivative of the distortion.

Vehkaoja *et al.* [58] introduced dynamic pressure sensors for detecting heartbeat intervals of an individual sleeping on a bed. The pressure sensors were composed of EMFi material and located under the bed supports. In this study, individual heartbeats were not observed. However, the intervals in which the correlation between two successive signals segment maximized. Ballistocardiogram signals were collected from nine subjects (5 males, 4 females) during 1-hour recording. The beat-to-beat intervals provided by this approach can be adopted in determining frequency domain heart rate variability that is most frequently used in the assessment of sleep quality.

Lee [59] *et al.* proposed to use load cells, installed under bed supports, to measure heart rate and respiration for infants. Four infants (5 to 42 months) were involved in the study and a total of 13 experiments were carried out between 10 to 178.8 minutes. Initially, heart rate and respiratory components were extracted using band-pass filters of various cutoff frequencies. For the heart rate component, a first-order differentiation filter was applied, thus a nonlinear transformation, i.e., a Shannon entropy was applied to the differentiated signal to obtain only positive peaks. Additionally, a moving average filter was employed to flatten out the spikes and noise bursts. At last, heart rate was measured by finding local peaks in an optimum signal. For the respiration component, as the band-pass filtered signal contained residual baseline drift, a detrending algorithm based on empirical mode decomposition was adopted to get rid of such unwanted trend. Similar to heart rate, local peaks were detected in the detrended signal and therefore the respiratory rate was measured. A signal quality index was developed to choose the optimum signal out of the four load cells' signals. The quality processing procedure was developed based on calculating a threshold value computed from an autocorrelation function and a power spectral density function. The proposed system achieved acceptable results compared to the reference ECG and respiratory belt. Table V summarizes the unconstrained monitoring of vital signs using the strain gauges-based sensors.

## VI. Hydraulic-Based Sensors

The concept of the hydraulic sensor is to measure the change in pressure applied to a liquid-filled tube. For example, Heise *et al.* [60] designed a hydraulic based-sensor for unrestrained monitoring of heart rate and respiration. Preliminary data were collected from two individuals (1 male and 1 female). Participants were instructed to lie on a bed for approximately 10 minutes. During the 10 minutes, they were asked to lie on the back, on the right side, on the back again, on the left side, and on the back once more (2 minutes each position). In this preliminary research, heartbeat signal was extracted by detecting the difference between the most negative and the most positive points within a moving window. After that, a low-pass filtered was applied to reduce the effect of noise and smooth the signal. A fixed threshold was employed to detect a body motion. Finally, the heart rate was measured by adopting the autocorrelation function. However, the respiratory rate was measured by low-pass filtering the signal and then subtracting the DC bias. Afterward, the zero-crossings were counted to provide the breaths per minute. Preliminary results approved that the hydroponic sensor was effective at extracting heart rate and respiration against the reference devices, namely a piezoresistive device worn on the subject's finger and respiration band wrapped around the subject's torso. In a different study, Heise *et al.* [61] have validated the sensor using data collected from five subjects (3 males and 2 females) and have confirmed stability of the signal processing algorithms using real and synthesized signals.

TABLE V
SUMMARY OF UNCONSTRAINED MONITORING OF VITAL SIGNS USING PNEUMATIC-BASED SENSORS. *N/A*: NOT AVAILABLE, *M*: MALE, *F*: FEMALE, *HR*: HEART RATE, *RR*: RESPIRATORY RATE, *Min*: MINUTES, *Hrs*: HOURS, *Sec*: SECONDS, *SWM/M*: SLIDING WINDOW MINIMUM/MAXIMUM, *ECG Sync*: ELECTROCARDIOGRAM SYNCHRONIZATION, *PCA*: PRINCIPAL COMPONENT ANALYSIS, *CCF*: CROSS-CORRELATION FUNCTION, *ED*: EUCLIDEAN DISTANCE, *HVS*: HEART VALVE SIGNAL, *ACF*: AUTOCORRELATION FUNCTION, *SE*: SHANNON ENTROPY, *EMD*: EMPIRICAL MODE DECOMPOSITION, *Lab*: LABORATORY.

| | Method | Subjects (M, F) | Deployment | Duration | Outcome |
|---|---|---|---|---|---|
| [51] | SWM/M | 2 M, 2 F | Lab | 5 Min | HR, RR |
| [52] | ECG Sync | 11 M, 10 F | Lab | 45 Sec | HR |
| [56] | PCA, K-means CCF, ED, HVS | 9 M, 7 F | Lab | 30 Min | HR |
| [58] | ACF | 5 M, 4 F | Lab | 1 Hrs | HR |
| [59] | SE, EMD, SWM/M | Infants (3 M, 1 F) | Home | 10 - 178.8 Min | HR, RR |

Rosales *et al.* [62] deployed four hydraulic transducers under the bed mattress, covering the upper part of the body in order to measure heart rate in a nonrestrictive way. Each transducer was connected to a pressure sensor to record the pressure forces applied to it. In this preliminary study, heartbeats were computed using a clustering-based approach as follows. Every five seconds, body motions were eliminated based on the variance of the transducers' signal. Following body motions removal, the transducer's signal was band-pass filtered to remove respiratory components and filtered once more using an average filter to smooth the signal prior to feature extraction. Afterward, three features were extracted from every 5-second time window based on the IJK points of the ballistocardiogram signal. In addition, the extracted features were classified into two groups using k-means clustering algorithm. The first group, i.e., the smallest cluster was assigned to the heartbeat class. Then, the second group, i.e., the largest cluster was assigned to the non-heartbeat group. In conclusion, the heartbeats' (J-peaks) locations were compared to a reference signal obtained from a piezoresistive device worn on the subject's finger. Data were acquired from four subjects (2 males and 2 females) during 6 minutes (supine position). Although such clustering-based approach might have provided good results it might only be applicable to specific situations. Furthermore, to think the presented method to be applied in practical applications, manually labeling (training) data is, however, a restricting property.

A similar study was proposed by Su *et al.* [63]. Nonetheless, the heart rate was measured using the Hilbert transform and the fast Fourier transform (30-second window). In this study, ballistocardiogram signals were acquired from five subjects (3 males and 2 females) during 2.5 minutes in a supine position. This approach provided a lower error rate compared with the windowed peak to peak deviation (WPPD) method introduced by Heise *et al.* [60]. Although results were consistent with the reference device, ballistocardiogram signals were assumed relatively stationary. This assumption is not always true because typically heartbeats are not uniform in time [64].

In another study, Lydon *et al.* [65] proposed a new algorithm to detect heart rate using the four hydraulic transducers. As a first step, a band-pass filter was implemented to remove the respiration component as well as high-frequency noise. Next, the data from the four transducers were separated into 0.3-second (30 samples) segments and the short-time energy profiles were computed for each segment. As a result, four hear rate values were generated for each transducer by locating the local peaks. Moreover, a single heart rate value was selected based on the DC level of each transducer's signal. Typically, a higher DC level in the obtained transducer's signal means that the transducer makes better contact with the body and therefore gives a more stable ballistocardiogram signal. Hence, the transducer with the highest DC level was chosen for heart rate measurement. Finally, outliers were eliminated by following whether the estimated heart rate value was more than 15 beats per minute from the moving average heart rate value. Validation data were collected from two groups, i.e, three subjects (2 males and 1 female) during 10 minutes recording and four older adults (4 males) in a typical home environment. This approach provided slightly better results compared to the clustering-based approach provided by Rosales *et al.* [62].

In order to address the uncertainty inherent in a ballistocardiogram signal, for instance, misalignment between training data and ground truth, improper collection of the heartbeat by some transducers, Jiao *et al.* [66] applied the Extended Function of Multiple Instances (eFUMI) algorithm to ballistocardiogram signals generated by the four hydraulic transducers. The objective of the eFUMI was to learn a personalized concept of heartbeat for a subject in addition to several non-heartbeat background concepts. Following the learning step, heartbeat detection and heart rate estimation can be applied to test data. The limitation of this algorithm is the need for sufficient training data, which might not be always available.

Rosales *et al.* [67] applied the clustering-based approach [62] and the Hilbert transform approach [63] to ballistocardiogram signal collected from four male senior residents. The signals were collected from residents over a two to four months period under in-home living conditions. However, the analysis was done only over five minutes of initial recordings. The Hilbert transform approach was able to produce more stable heart rate estimates compared to the clustering-based approach. The latter approach was more susceptible to motion artifacts. Table VI summarizes the unconstrained monitoring of vital signs using the hydraulic-based sensors.

## VII. Fiber Optic-Based Sensors

In existing literature, unobtrusive vital signs monitoring is achieved either by microbend fiber-optic sensors (MFOS) or fiber Bragg grating sensors (FBGS). The principle of the MFOS is that if an optical fiber is bent, insignificant amounts of light are lost through the fiber walls. This reduces the amount of received light and is a function of bend pressure [68], [69], [70], [71]. The FBG is an optical fiber that serves as a filter for a specific wavelength of light. The principle of the FBGS is to detect the reflected Bragg wavelength shift owing to changes in temperature, strain, or pressure [72], [73]. Fiber Bragg gratings are commonly used optical fiber sensors for measuring temperature and/or mechanical strain. Though, the excessive cost of the interrogation systems is the most significant obstacle for their large commercial application [74].

Chen *et al.* [75], [76] described the effectiveness of the MFOS for nonintrusive monitoring of heart rate and breathing rate. For heart rate, ballistocardiogram signals were gathered from several subjects in sitting position and breathing normally. Preliminary results have proved that the ballistocardiogram waveforms closely simulated those reported in the existing literature. For breathing rate, nine volunteers were involved in the study in which respiratory signals were collected during sleep. The system has shown a good match with the reference respiratory device. Deepu *et al.* [77] introduced a smart cushion integrated with MFOS for real-time heart rate monitoring. The cushion can be placed on the seat or back of a chair for data collection. In this study, five subjects were involved, and signals were collected during 5-minutes. Several steps were applied to the cushion's signals in order to unobtrusively measure the heart rate. Initially, low and high-frequency noises were suppressed using a band-pass finite impulse response (FIR) filter. Next, a cubing operation was applied to the filtered signal to enhance the amplitude swing while keeping the signal sign intact. Afterward, momentary upswing or downswing was removed by applying a moving average filter. Furthermore, the resultant signal was smoothed by utilizing the absolute value and averaging over a predefined time window. At last, the J-peaks were recognized by using a cone detection and comparing to an adaptive threshold. The proposed system achieved satisfactory results compared to the reference pulse oximetry device.

TABLE VI
SUMMARY OF UNCONSTRAINED MONITORING OF VITAL SIGNS USING HYDRAULIC-BASED SENSORS. *N/A*: NOT AVAILABLE, *M*: MALE, *F*: FEMALE, *HR*: HEART RATE, *RR*: RESPIRATORY RATE, *Min*: MINUTES, *Hrs*: HOURS, *Sec*: SECONDS, *WPPD*: WINDOWED PEAK TO PEAK DEVIATION, *CA*: CLUSTERING APPROACH, *HT*: HILBERT TRANSFORM, *STE*: SHORT-TIME ENERGY, *eFUMI*: EXTENDED FUNCTION OF MULTIPLE INSTANCES, *Lab*: LABORATORY.

|  | Method | Subjects (M, F) | Deployment | Duration | Outcome |
|---|---|---|---|---|---|
| [60] | WPPD | 1 M, 1 F | Lab | 10 Min | HR, RR |
| [61] | WPPD | 3 M, 2 F | Lab | 10 Min | HR, RR |
| [62] | CA | 2 M, 2 F | Lab | 6 Min | HR |
| [63] | HT | 3 M, 2 F | Lab | 2.5 Min | HR |
| [65] | STE | 2 M, 1F | Lab | 10 Min | HR |
|  |  | 4M | Home | Overnight | HR |
| [66] | eFUMI | 4 N/A | Lab | 10 Min | HR |
| [67] | CA, HT | 4 M | Home | Overnight | HR |

Chen *et al*. [78] studied the possibility of measuring blood pressure using ballistocardiography and photoplethysmography (PPG). The concept was to calculate the time delay between the peaks of the ballistocardiography and the corresponding PPG peaks. Ballistocardiogram signals were collected from five healthy subjects in a sitting position using a cushion integrated with MFOS, whereas PPG signals were collected from a finger pulse oximeter. Preliminary results have shown that blood pressure might be measured using optical devices. However, the proposed approach was very challenging because it required a calibration procedure for each subject prior to measurement.

Lau *et al*. [79] evaluated the effectiveness of the MFOS for respiratory monitoring and respiratory gating in the magnetic resonance imaging (MRI) environment. Respiratory gating is the process of reducing cardiorespiratory artifacts by synchronizing magnetic resonance data acquisition to the cardiac or respiratory cycles. Unlike electrical sensors, fiber-optic sensors are immune to electromagnetic and radio-frequency interference. Twenty healthy subjects (10 males and 10 females) were involved in the study and they underwent T2-weighted half-Fourier single-shot turbo spin-echo MRI of the liver with synchronous breathing rate monitoring on a 1.5 Tesla magnetic resonance scanner. The breathing rate was detected by applying a band-pass filter and hence detecting local peaks in the time domain. This study presented that the MFOS were able to detect comparable breathing rate to the reference respiratory bellows and produce liver MRI images of good diagnostic quality compared to the navigator-acquired scans. Chen *et al*. [80] reported related results using data collected from eleven healthy subjects (6 males and 5 females) during MRI.

A similar study was provided by Dziuda *et al*. [81]. However, authors used FBG sensors rather than MFOS. Three healthy volunteers (2 males and 1 female) were included in the study and physiological data were collected during 95 minutes. Both heart rate and breathing rate were measured by finding local maxima after applying band-pass filters of different cutoff frequencies to the sensor data. Similar to the MFOS, the FBG sensor did not introduce any artifacts into MRI images. Furthermore, the system achieved comparable results to the reference devices, i.e., carbon electrodes and pneumatic bellows, respectively. Dziuda *et al*. [82], [83], [84], [85] reported similar results using data collected during MRI examination.

Zhu *et al*. [86] demonstrated the effectiveness of the MFOS for unobtrusive measurement of heart rate in a headrest position. Three healthy individuals were enrolled in the study in which an optical sensor mat was placed on the headrest of a massage chair. The participants were instructed to complete predefined series of tasks, i.e., rest, cognitive test battery, and relaxing massage session. In this preliminary study, the analysis was done only during rest periods for a total of six minutes. A band-pass filter was applied to the sensor data to remove low-frequency respiratory signals. Afterward, heart rate was computed using short-time Fourier transform. The proposed system achieved a relatively good agreement against the reference ECG.

Chen *et al*. [87] reported the results of using the MFOS in a clinical trial for unobtrusive monitoring of heart rate and respiration during sleep. During the study, data were collected from twenty-two subjects using the optical fiber sensor and also from the standard polysomnography as a reference. In the beginning, large body movements were eliminated using a moving time window. In which, a segment was identified as a body movement if the difference between the maximum and the minimum in the moving window was larger than a fixed threshold. Next, respiratory and heartbeat components were separated from the sensor's signals using band-pass filters of different cutoff frequencies. In the former, the signals were smoothed using a moving-average filter and hence the baseline was obtained by another moving-average filter of a larger window size. After subtracting the signals and the baseline, they were further smoothed using Savitzky–Golay method. Finally, local peaks were detected, and breathing rate was computed. In the latter, all local peaks of the heartbeat signals were detected, and heart rate was computed accordingly. Consequently, incorrect heart rate values were eliminated by applying a histogram-based method, in which the group with the highest occurrence was selected and reported as final heart rate results. Results were promising. However, the proposed approach was prone to motion artifacts.

Zhu *et al*. [88] proposed to measure heart rate using ballistocardiogram signals collected from FBG sensor mat. The sensor mat consisted of three FBG sensor arrays or channels and each array contained six sensors. The arrays were located under the pillow, upper chest, and lower chest. In this study, ten subjects were enrolled, and signals were collected during 20 minutes such as 10 minutes of supine posture and 10 minutes of sideways posture. ECG signals were collected along with the fiber-optic signals as a reference. The signal from each sensor array was transformed from time domain into cepstrum domain. After that, the signal from the six sensors of the same arrays was fused by employing cepstrum. Finally, the heart rate was measured from the fused signal by recognizing peaks in the cepstrum. This study demonstrated that the heart rate can be measured from distinct locations. However, the best results were achieved from sensor arrays at chest position. In another study, Zhu *et al*. [89] used the same system to compute breathing rate and the system was tested against twelve subjects.

Fajkus *et al*. [90] introduced to measure heart rate and respiration using FBG sensors encapsulated inside a polydimethylsiloxane polymer (PDMS). The FBG sensors were embedded within a thoracic elastic strap to record cardiorespiratory signals. In this preliminary analysis, the authors collected data from 10 individuals (6 males and 4 females) during few minutes.

Heart rate and breathing rate were detected by adopting two methods, i.e., identifying the periodic cycles in the time domain and applying the FFT to obtain the dominant frequency. The proposed system achieved comparable results to the reference ECG. However, it was susceptible to large body movements. In another study, Fajkus *et al*. [91] assessed the effectiveness of using FBG sensor encapsulated inside a PDMS and FBG sensor glued on a plexiglass pad for heart and respiratory rate monitoring. In this preliminary study, the authors collected data from 10 subjects (7 males and 3 females) and result shown that the FBG sensor encapsulated into PDMS was more accurate than FBG sensor encapsulated in plexiglass pad.

Chethana *et al*. [92] reported the use of FBG sensor for monitoring cardiac and breathing activities. Cardiorespiratory signals were collected from four subjects (2 males and 2 females) for 60 seconds, on which the FBG sensor was placed on the pulmonic area on the chest of the subjects. Results have been evaluated against an electronic stethoscope which recognizes and records sound pulses generated from the cardiac activity. Nedoma *et al*. [93] evaluated the effectiveness of the FBG sensor against fiber interferometric sensor for heart rate measurement. The former measured the heart rate through ballistocardiography, while the latter measured the heart rate through Phonocardiography. Cardiac signals were obtained from six individuals (3 males and 3 females) using the two sensors for 60 minutes. Primary results have shown that the fiber interferometric sensor was more accurate than the FBG sensor. Table VII summarizes the unconstrained monitoring of vital signs using the fiber optic-based sensors.

## VIII. CONCLUSION

This Paper provided the definition and the nomenclature of ballistocardiography. In addition, it discussed in detail the different modalities reported in existing literature for unobtrusive monitoring of vital signs, namely heart rate, breathing rate, and body movements. These modalities include piezoelectric polyvinylidene fluoride sensors, electromechanical film sensors, pneumatic sensors, load cells, hydraulic sensors, and fiber-optic sensors. In general, the output of these sensors is a composite signal that is composed of cardiac activities, respiratory activities, and body movements. Hence, these three signals should be separated from each other so that vital signs can be measured. The separation process is usually performed by applying a band-pass filter of specific cutoff frequencies according to the signal of interest. In other cases, the separation process can be performed by adopting a decomposition algorithm such as empirical mode decomposition algorithm and wavelet multiresolution analysis. It should be noted that, vital activities cannot be detected during body movements and hence they should be eliminated prior to the measurement process. Following the separation process, i.e., obtaining cardiac signals and respiratory signals, several algorithms can then be implemented for vitals measurements. As discussed in previous sections, these algorithms include but not limited to simple peak detector, autocorrelation function, fast Fourier transform, cepstrum analysis, wavelet multiresolution analysis, empirical mode decomposition, power spectrum analysis, and clustering-based approaches. The clustering-based approaches are not very effective because the training step should be repeated whenever the data collection protocol has been changed. Moreover, the ballistocardiogram morphology varies between and within subjects, and the shape of the signal is highly dependent on subject's postures, i.e., sleeping or sitting. Furthermore, the raw signal is noisy and nonstationary due to body movement, induced respiratory efforts, and the characteristics of the sensing system itself.

TABLE VII
SUMMARY OF UNCONSTRAINED MONITORING OF VITAL SIGNS USING HYDRAULIC-BASED SENSORS. *N/A*: NOT AVAILABLE, *M*: MALE, *F*: FEMALE, *HR*: HEART RATE, *RR*: RESPIRATORY RATE, *Min*: MINUTES, *Hrs*: HOURS, *Sec*: SECONDS, *BP*: BLOOD PRESSURE, *PPG Sync*: PHOTOPLETHYSMOGRAPHY SYNCHRONIZATION; *STFT*: SHORT-TIME FOURIER TRANSFORM, *CEPS*: CEPSTRUM, *Lab*: LABORATORY.

|  | Method | Subjects (M, F) | Deployment | Duration | Outcome |
|---|---|---|---|---|---|
| [75] | Visually | N/A | Lab | N/A | HR |
| [76] | Visually | 9 N/A | Lab | N/A | RR |
| [77] | Peak Detector | 5 N/A | Lab | 5 Min | HR |
| [78] | PPG Sync | 5 N/A | Lab | N/A | BR |
| [79] | Peak Detector | 10 M, 10 F | MRI | N/A | RR |
| [80] | Peak Detector | 6 M, 5 F | MRI | N/A | HR, RR |
| [81] | Peak Detector | 2 M, 1 F | MRI | 95 Min | HR, RR |
| [82] | Peak Detector | 8 M, 4 F | MRI | 60 Min | HR, RR |
| [83] | Peak Detector | 1 M | MRI | 19 Min | HR |
| [84] | Peak Detector | 6 M, 2 F | MRI | 82 Min | HR |
| [86] | STFT | 3 N/A | Lab | 6 Min | HR |
| [87] | Peak Detector | 22 N/A | Hospital | Overnight | HR, RR |
| [88], [89] | CEPS | 10 N/A | Lab | 20 Min | HR, RR |
| [90] | Peak Detector, FFT | 6 M, 4 F | Lab | N/A | HR, RR |
| [92] | Visually | 2 M, 2 F | Hospital | 1 Min | HR, RR |
| [93] | Peak Detector | 3 M, 3 F | Lab | 60 Min | HR, RR |